\documentclass{ptapap}

\author{Lorenzo Rimoldini}[UNIGE-ECO]
\author{Laurent Eyer}[UNIGE] 
\author{Nami Mowlavi}[UNIGE]
\author{Dafydd W. Evans}[UCAM]
\author{Krzysztof Nienartowicz}[SIXSQ]
\author{Berry Holl}[UNIGE-ECO] 
\author{Marc Audard}[UNIGE-ECO]
\author{Leanne P. Guy}[UNIGE-ECO]
\author{Gr\'egory Jevardat de Fombelle}[SIXSQ]
\author{Isabelle Lecoeur-Ta\"ibi}[UNIGE-ECO]
\author{Olivier Marchal}[UNIGE-ECO]
\author{Gisella Clementini}[OAB] 
\author{Vincenzo Ripepi}[OAC]
\author{Alessia Garofalo}[UB,OAB]
\author{Roberto Molinaro}[OAC]
\author{Tatiana Muraveva}[OAB]
\author{Ennio Poretti}[OB]
\author{L\'aszl\'o Moln\'ar}[KO]
\author{Emese Plachy}[KO]
\author{\'Aron Juh\'asz}[KO,EU]
\author{L\'aszl\'o Szabados}[KO]
\author{Joris De Ridder}[KU]
\author{Sara Regibo}[KU]
\author{Luis Manuel Sarro Baro}[UNED]
\author{Mauro L\'opez del Fresno}[CSIC]

\affil[UNIGE-ECO]{Department of Astronomy, University of Geneva, Chemin d'Ecogia 16, \\CH-1290 Versoix, Switzerland}
\affil[UNIGE]{Department of Astronomy, University of Geneva, Chemin des Maillettes 51, \\CH-1290 Versoix, Switzerland}
\affil[UCAM]{Institute of Astronomy, University of Cambridge, Madingley Road, \\Cambridge CB3 0HA, United Kingdom}
\affil[SIXSQ]{SixSq, Rue du Bois-du-Lan 8, CH-1217 Meyrin, Switzerland}
\affil[OAB]{INAF - Osservatorio Astronomico di Bologna, Via Gobetti 93/3, \\I-40129 Bologna, Italy}
\affil[OAC]{INAF - Osservatorio Astronomico di Capodimonte, Via Moiariello 16, \\I-80131 Napoli, Italy}
\affil[UB]{Department of Physics and Astronomy, University of Bologna, Via Gobetti 93/2, \\I-40129 Bologna, Italy }
\affil[OB]{INAF - Osservatorio Astronomico di Brera, Via E. Bianchi 46, I-23807 Merate, Italy}
\affil[KO]{Konkoly Observatory, Research Centre for Astronomy \& Earth Sciences, Hungarian Academy of Sciences, Konkoly Thege Mikl\'os \'ut 15-17, H-1121 Budapest, Hungary}
\affil[EU]{E\"otv\"os Lor\'and University, Egyetem t\'er 1-3, H-1053 Budapest, Hungary}
\affil[KU]{Institute of Astronomy, KU Leuven, Celestijnenlaan 200D, B-3001 Leuven, Belgium}
\affil[UNED]{Departamento Inteligencia Artificial, UNED, Calle Juan del Rosal 16, \\E-28040 Madrid, Spain}
\affil[CSIC]{Departamento de Astrof\'isica, Centro de Astrobiolog\'ia (INTA-CSIC), PO Box 78, \\E-28691 Villanueva de la Ca\~nada, Spain}

\title{All-sky RR~Lyrae Stars in the \textit{Gaia} Data}

\begin{document}

\maketitle

\begin{abstract}

The second \textit{Gaia} data release is expected to contain data products from about 22 months of observation. Based on these data, we aim to provide an advance publication of a full-sky \textit{Gaia} map of RR~Lyrae stars. 
Although comprehensive, these data still contain a significant fraction of sources which are insufficiently sampled for Fourier series decomposition of the periodic light variations. The challenges in the identification of RR~Lyrae candidates with (much) fewer than 20 field-of-view transits are described. 
General considerations of the results, their limitations, and interpretation are presented together with prospects for improvement in subsequent \textit{Gaia} data releases.

\end{abstract}

\section{Introduction}

RR~Lyrae stars are particularly useful variable objects as they combine ease of detection (due to light variations of up to about two magnitudes and periods typically less than about one day) with the benefits of being standard candles, making studies of structures and  distance estimations possible within and beyond our Galaxy.

A low number of observations can suffice in the identification of RR~Lyrae stars \citep{2000AJ....120..963I,2016ApJ...817...73H}, so
we searched for these objects in the first 22 months of data from the \textit{Gaia} mission \citep{2016A&A...595A...1G}, with the hope that 
the community will benefit of an advance publication of all-sky RR~Lyrae candidates in \textit{Gaia} Data Release~2 (DR2).
Efforts to identify RR~Lyrae stars in \textit{Gaia}~DR1 data from sources without time series have already appeared in the literature \citep[e.g.,][]{2017MNRAS.466.4711B,2018MNRAS.474.2142I}, indicating that an early publication and exploitation of selected \textit{Gaia} time series can better address current challenges. 
Herein, the term `observation' denotes a single field-of-view transit of a source along the \textit{Gaia} focal plane in the $G$~band (typically a combination of 8 or 9 astrometric field CCDs).

\section{Method}

The minimum number of observations necessary for reliable Fourier series decomposition of the periodic light variations depends on several factors, like the time sampling, the range of fundamental frequencies of the light curve, the number of harmonics needed to characterize the time series, and the level of noise. In the case of \textit{Gaia}~DR1, at least 20 observations were required for most of the published RR~Lyrae stars (only in a few cases the number of observations was as low as $12-15$).
For \textit{Gaia}~DR2 candidates, more than half of the objects have less than 20 observations, with the highest peak of the distribution around $13-14$ observations.

In order to recognize RR~Lyrae stars, machine-learning models are built with feature-based semi-supervised classification techniques. 
For an unbiased identification of candidates associated with high and low (at least two) numbers of observations,
classifiers are trained with values of statistical nature without resorting to Fourier modeling parameters.  
The classification of sources with at least 20 observations, covering (non-uniformly) about half of the sky and including Fourier parameters, is performed by an independent pipeline run, and the results from both approaches will be published in \textit{Gaia}~DR2.

\paragraph{Crossmatch with Literature.}

The crossmatch of \textit{Gaia} sources with objects from the literature is fundamental to the construction of a realistic training set and the subsequent validation of results. Known RR~Lyrae stars are extracted from the following surveys:
ASAS \citep{1997AcA....47..467P},
Catalina \citep{2009ApJ...696..870D},
\textit{Gaia} \citep{2016A&A...595A...2G,2017arXiv170203295E,2016A&A...595A.133C},
\textit{Hipparcos} \citep{HipparcosESA,Perryman1997Hipparcos},
LINEAR \citep{2000Icar..148...21S},
NSVS \citep{2004AJ....127.2436W},
OGLE-IV \citep{2015AcA....65....1U},
Pan-STARRS1 \citep{2016arXiv161205560C}, and
SDSS \citep{2009ApJS..182..543A},
in addition to a selection of globular clusters and ultra-faint dwarf spheroidal galaxies.

\paragraph{Classification Attributes.}

About 150 attributes are defined to characterize different aspects of the time series and a subset of 40 attributes is selected according to the attribute usefulness as perceived by the classifier \citep[e.g.,][]{Guyon.Elisseeff.Variable.Selection}.
The classification attributes that are found particularly relevant to the identification of RR~Lyrae stars in the \textit{Gaia} data include 
(i)~the amplitude of light variations, 
(ii)~the skewness of the distribution of magnitudes, 
(iii)~the typical magnitude range (and its fraction with respect to the full time series) when restricted to time series segments of one or half a day, and 
(iv)~the interquartile range of the distribution of absolute values of magnitude changes per unit time between successive observations.

\paragraph{Classification Model.} 

Crossmatched objects 
unavoidably suffer from the selection biases present in the originating catalogs, due to the peculiarities and limitations of each survey (sky coverage, sampling, photometric bands, sensitivity, etc.). Resampling is used to alleviate density peaks in the distributions of some parameters, 
and a semi-supervised approach is applied to the training set of certain classes (constant stars, fundamental and first-overtone RR~Lyrae stars, and Mira variables) by selecting results from a previous classification run to  
improve the training set representation in the sky and in magnitude distribution.
The selection of unlabeled sources to add to the training set depends on the parameter gap(s) to fill and on the classification probability (sufficiently high to limit chances of contamination, but not too high in order to allow the discovery of new information).
In order to ensure confidence in the new objects, unlabeled sources are verified through basic statistical filters as a function of variability type before adding them to the training set.

Random Forest \citep{Breiman.Random.Forest} classifiers are implemented in a multi-stage setting which combines a set of dedicated classifiers to solve simpler problems with fewer attributes, making data less diluted and thus better represented in attribute space. 
Multi-stage classification avoids also the potentially counter-productive competition which could occur in the identification of major classes and sub-classes at the same level. 
Our multi-stage classifier includes five dedicated classifiers to separate 
(i)~constant objects from low-amplitude variables and other variables,
(ii)~low-amplitude variability types,
(iii)~other variability classes,
(iv)~RR~Lyrae stars into sub-types (fundamental, first overtone, double mode and anomalous double mode), and
(v)~Cepheids into sub-types (anomalous, classical, type-II).
Classifications from (ii) and (iii) depend on the results of (i), while those from (iv) and (v) depend on~(iii).

\begin{figure}[ht]
\center
\includegraphics[width=0.64\textwidth]{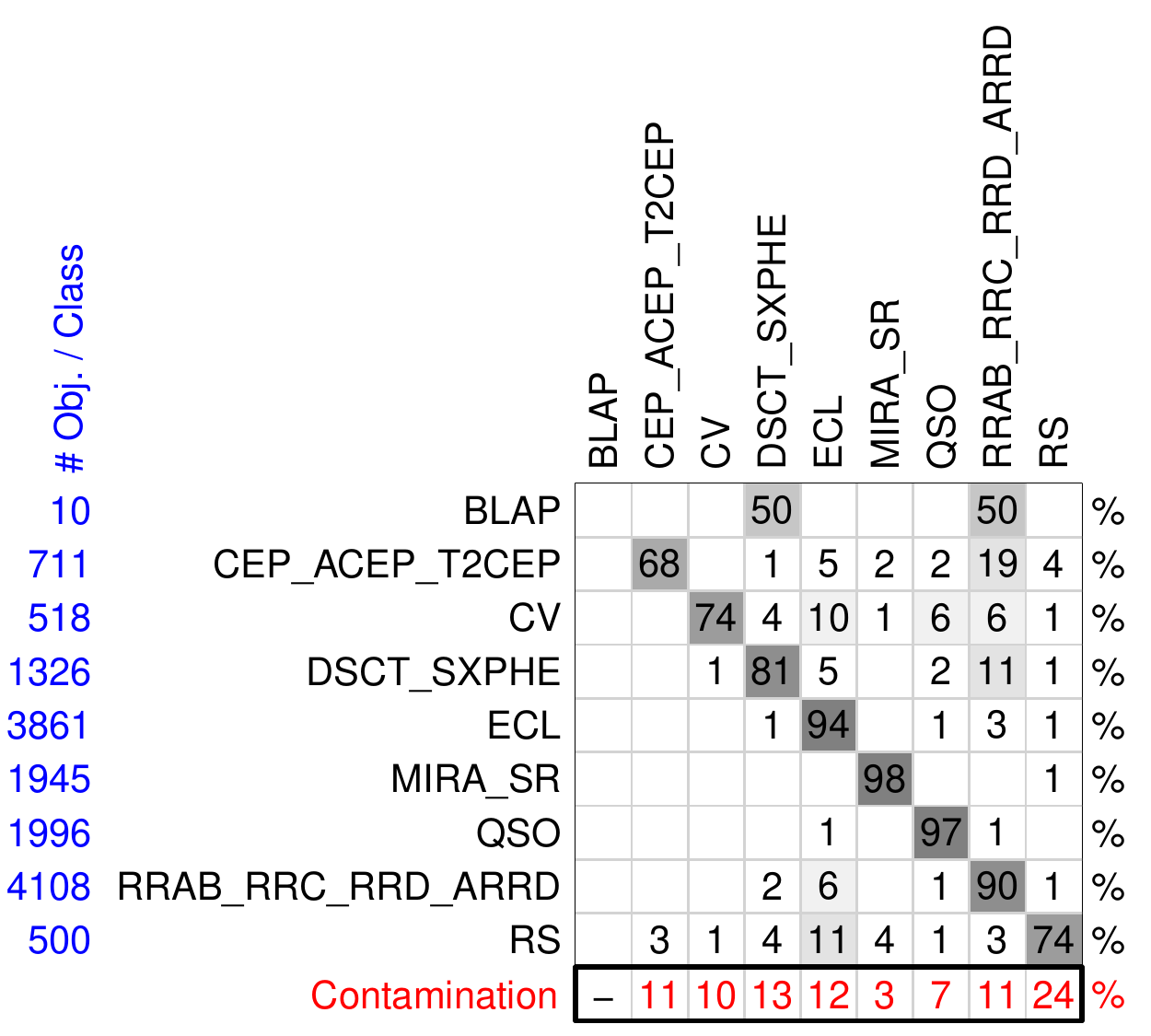}
\caption{Confusion matrix of the classifier addressing the main classes published in \textit{Gaia}~DR2: training-set objects (in rows) and their numbers (on the left-hand side) versus predictions from the classifier (in columns), where percentages along the diagonal represent the  completeness rates. Trained classes include BLAP (blue large amplitude pulsators), CEP\_ACEP\_T2CEP (classical, anomalous, and type-II Cepheids), CV (cataclysmic variables), DSCT\_SXPHE ($\delta$~Scuti and SX~Phoenicis stars), ECL (eclipsing binaries), MIRA\_SR (Mira and semi-regular variables), QSO (quasars), RRAB\_RRC\_RRD\_ARRD (fundamental, first-overtone, double-mode, and anomalous double-mode RR~Lyrae stars), and RS (RS~Canum Venaticorum-type binary systems).  Candidates of the under-represented BLAP class are merged with those of the DSCT\_SXPHE class in \textit{Gaia}~DR2.}
\label{fig:confusion_matrix}
\end{figure}

A few other variability types (besides RR~Lyrae stars) from the all-sky classification will also be published in \textit{Gaia}~DR2: Cepheids, high-amplitude $\delta$~Scuti and SX~Phoenicis stars, and long period (Mira and semi-regular) variables. Additional classes are included (although not published) in different classification stages to reduce the contamination of the targeted classes.
The average completeness and contamination rates (per type) of the classifier that distinguishes the main variability types published in \textit{Gaia}~DR2 are shown in Fig.~\ref{fig:confusion_matrix} and are applicable to sources with attribute distributions similar to those of the training set.

\section{Results}

The classification results are associated with a score to quantify the reliability of the candidates.
The number of true and false positives among the RR~Lyrae classifications is assessed statistically in terms of completeness and contamination rates as a function of classification score, magnitude, and number of observations. 
These rates vary also as a function of extinction, reddening, crowding, and possibly other parameters, which need to be considered (depending on the context of the analysis) for a correct interpretation of the identified candidates. 

As expected, the distribution in the sky reveals the absence of reliable candidates in the most reddened and extinguished regions of the Galactic disc.
Many known objects such as globular clusters, dwarf spheroidal galaxies, and the Sagittarius tidal streams are easily distinguishable in a simple sky map of RR~Lyrae candidates.

Multiple independent validations of the results are performed for each of the published variability types. In the case of RR~Lyrae classifications, such validations include:
(i)~known objects from the literature to assess the statistical quality of results and reduce obvious contaminants;
(ii)~sources detected in the \textit{Kepler}/\textit{K2} fields (characterized by finely sampled light curves) to search for counterparts of \textit{Gaia} candidates and estimate the fraction of missed identifications;
(iii)~the pipeline module dedicated to the confirmation of RR~Lyrae stars, 
usually executed with at least 20 observations, but now extended  to include sources with as few as 12 observations.
Full details of the results and comparisons with recent works in the literature will be presented in the documentation and articles accompanying \textit{Gaia}~DR2.

\section{Prospects}

The data in \textit{Gaia}~DR2 will include astrometric information, time series in the $G$, $G_{\rm BP}$, $G_{\rm RP}$ bands, and statistical parameters of the published RR~Lyrae candidates. More detailed information (such as period, modeling parameters, metallicity and extinction) will be available for a subset of the sources 
with at least 12 observations
that could be confirmed by the RR~Lyrae verification module with detailed Fourier modeling.
Future \textit{Gaia} data releases are expected to provide identifications of RR~Lyrae stars with higher completeness and lower contamination rates, due to more observations, improved astrometry, radial velocities, spectra, and astrophysical parameters (such as temperature and extinction), some of which might already be available in \textit{Gaia}~DR2 (although not in time to be used in this classification).

\acknowledgements{This work made use of data from the ESA space mission \textit{Gaia}, processed by the \textit{Gaia} Data Processing and Analysis Consortium. Support included the Bolyai Research Scholarship of the HAS (L.M., E.Pl.), \'UNKP-17-3 program of the Ministry of Human Capacities of Hungary (\'A.J.), and NKFIH grants PD-116175, PD-121203, K-115709.}

\bibliographystyle{ptapap}
\bibliography{RRL2017_rimoldini}

\end{document}